\documentclass[preprint,journal]{vgtc}            

\onlineid{0}

\preprinttext{}

\vgtccategory{Research}

\title{NeuVolEx: Implicit Neural Features for Volume Exploration}

\author{%
  Haill An$^{*}$, Suhyeon Kim$^{*}$,
  Donghyuk Choo, and 
  Younhyun Jung$^{\dagger}$
}

\authorfooter{
  \item $^{*}$ H. An and S. Kim contributed equally to this work.
  \item $^{\dagger}$ Corresponding author.
  \item H. An, S. Kim, D. Choo, and Y. Jung are with the School of Computing, Gachon University, Republic of Korea. 
        E-mail: \{xenotic, kih629, choodh105, younhyun.jung\}@gachon.ac.kr.
}

\abstract{%
    Direct volume rendering (DVR) aims to help users identify and examine regions of interest (ROIs) within volumetric data, and feature representations that support effective ROI classification and clustering play a fundamental role in volume exploration. Existing approaches typically rely on either explicit local feature representations or implicit convolutional feature representations learned from raw volumes. However, explicit local feature representations are limited in capturing broader geometric patterns and spatial correlations, while implicit convolutional feature representations do not necessarily ensure robust performance in practice, where user supervision is typically limited. Meanwhile, implicit neural representations (INRs) have recently shown strong promise in DVR for volume compression, owing to their ability to compactly parameterize continuous volumetric fields. In this work, we propose NeuVolEx, a neural volume exploration approach that extends the role of INRs beyond volume compression. Unlike prior compression methods that focus on INR outputs, NeuVolEx leverages feature representations learned during INR training as a robust basis for volume exploration. To better adapt these feature representations to exploration tasks, we augment a base INR with a structural encoder and a multi-task learning scheme that improve spatial coherence for ROI characterization. We validate NeuVolEx on two fundamental volume exploration tasks: image-based transfer function (TF) design and viewpoint recommendation. NeuVolEx enables accurate ROI classification under limited user supervision for image-based TF design and supports unsupervised clustering to identify compact complementary viewpoints that reveal different ROI clusters. Experiments on diverse volume datasets with varying modalities and ROI complexities demonstrate NeuVolEx consistently improves both effectiveness and usability over prior methods.
}


\teaser{
  \centering
  \includegraphics[width=\linewidth, alt={A view of a city with buildings peeking out of the clouds.}]{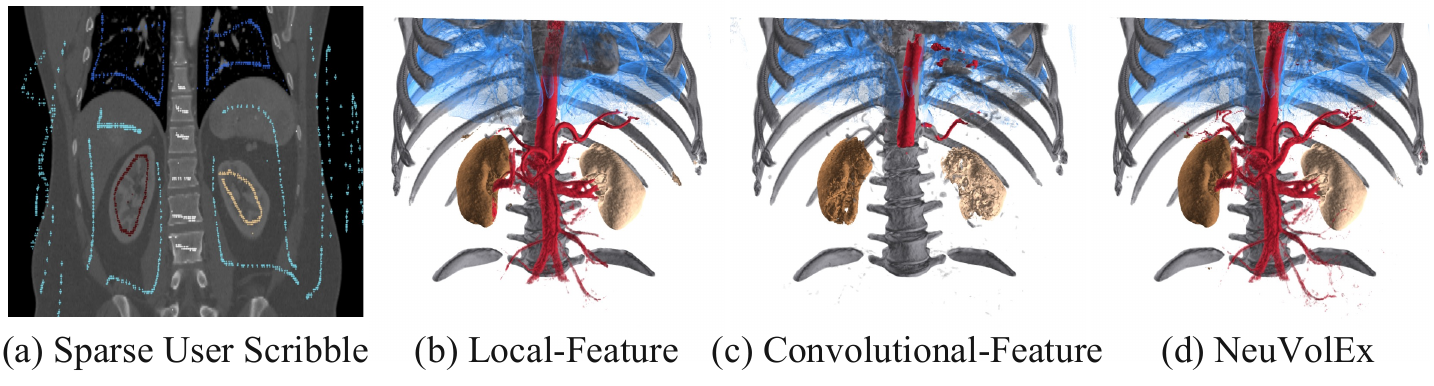}
  \caption{%
    Image-based TF design comparison among NeuVolEx, the explicit local-feature method \cite{soundararajan2015learning}, and the implicit convolutional-feature method \cite{quan2017intelligent} based on sparse user scribbles on CT-Abdomen.
  }
  \label{fig:teaser}
}

\graphicspath{{figs/}{figures/}{pictures/}{images/}{./}} 

\usepackage{tabu}                      
\usepackage{booktabs}                  
\usepackage{multirow}
\usepackage{lipsum}                    
\usepackage{mwe}                       
\usepackage{ccicons}                   
\usepackage{graphicx} 
\usepackage{makecell}
\usepackage{mathptmx}                  

\begin{document}


\firstsection{Introduction}

\maketitle


Direct volume rendering (DVR) is an indispensable tool for visualizing and interpreting volumetric data and has been widely applied in various domains, ranging from medical imaging to scientific simulations. A long-standing goal in the DVR community is to help users identify and examine regions of interest (ROIs) within volumes, which has motivated a variety of volume exploration techniques. Among them, transfer function (TF) design is one of the most widely used, which enables users to interactively classify and emphasize ROIs by mapping optical properties, such as opacity and color, to voxels. Another important direction is viewpoint recommendation, which supports initial exploration and rapid browsing by suggesting a set of complementary viewpoints that collectively reveal different ROIs. Although these tasks differ in how exploration is performed, both rely fundamentally on a feature representation that effectively characterizes ROIs; therefore, feature representation plays a central role in volume exploration. 

Feature representations for volume exploration are typically derived from raw volumes. One line of work explicitly extracts local voxel features, such as intensity, gradient magnitude, spatial location, texture, and size \cite{max2002optical, cai1995rendering, caban2008texture, correa2008size}. Although these explicit local features can faithfully reflect the intrinsic characteristics of voxels, their limited ability to encode broader geometrical patterns or spatial correlations often hinders coherent ROI classification and clustering. Another line of work uses deep neural networks to learn implicit hierarchical convolutional features from raw volumes. These learned representations can capture spatial context and global relationships beyond explicit local feature descriptors, but this advantage does not necessarily ensure robust performance in practice. In particular, the high dimensionality of implicit convolutional feature representations (e.g., 75D \cite{quan2017intelligent}, 200D \cite{cheng2018deep}, or 387D \cite{engel2024leveraging}) makes accurate ROI classification difficult under realistic interaction budgets, where user supervision is typically limited. In image-based TF design, for example, users generally provide only sparse scribble annotations on a few image slices, which makes high-dimensional representations difficult to use effectively. Their high dimensionality also incurs substantial computational and memory overhead during feature training. Consequently, existing approaches either lack coherent ROI characterization or struggle to achieve accurate results in practical volume exploration settings. 

Recently, implicit neural representations (INRs) have attracted growing attention in the DVR community. INRs have demonstrated strong potential for functional approximation of continuous signals or fields. In the DVR community, they have been actively explored for volume compression owing to their capability to compactly and implicitly parameterize continuous volumetric fields. Such methods learn a direct mapping from voxel coordinates to intensity values using a multi-layer perceptron (MLP), with the trained network serving as a compressed version of the original volume. This enables intensity queries at arbitrary coordinates without explicit decompression or interpolation, thereby enabling volume reconstructions at varying resolutions while maintaining voxel-specific characteristics. Furthermore, trainable input encodings, such as hash grids, enable even compact MLPs to capture complex volumetric fields with high training efficiency and low computational cost.   

In this work, we propose a novel volume exploration approach that extends the role of INRs beyond volume compression. Rather than focusing on INR outputs, our neural volume exploration (NeuVolEx) leverages feature representations learned during INR training as a robust basis for volume exploration. Although base INRs provide informative features for volumes, they are not inherently optimized for volume exploration tasks. Base INRs are primarily trained to approximate individual voxels accurately, whereas volume exploration requires spatially coherent grouping of adjacent voxels with subtle variations and separation from different local neighborhoods. To better encode this spatial coherence, NeuVolEx introduces a structural encoder and a multi-task learning scheme. We validate NeuVolEx on two representative volume exploration tasks: image-based TF design and viewpoint recommendation. Image-based TF design requires feature representations that enable accurate ROI classification under limited user supervision (as exemplified in \cref{fig:teaser}), whereas viewpoint recommendation requires unsupervised clustering of feature representations to identify a compact set of complementary viewpoints that collectively reveal different ROI clusters. We quantitatively and qualitatively evaluate NeuVolEx on a variety of volume datasets with different imaging modalities and ROI complexities and further assess its usability in terms of interaction efficiency, training efforts, and computational overhead. The main contributions of this work are summarized as follows: 

\begin{itemize}
    \item \textbf{Novel INR-based feature representation for volume exploration:} We introduce an INR-based feature representation as a robust basis for volume exploration.
    \item \textbf{Extension of the role of INR in DVR:} We extend INR beyond its conventional use in volume compression to enable volume exploration.
    \item \textbf{A volume exploration-optimized INR:} We augment a base INR with a structural encoder and a multi-task learning scheme to better encode spatial coherence for ROI characterization.
    \item \textbf{Comprehensive validation on two representative volume exploration tasks:} We demonstrate the effectiveness and usability of NeuVolEx across a variety of volume datasets as well as diverse task-specific settings.
\end{itemize}

\begin{figure*}[ht]
   \includegraphics[width=\textwidth]{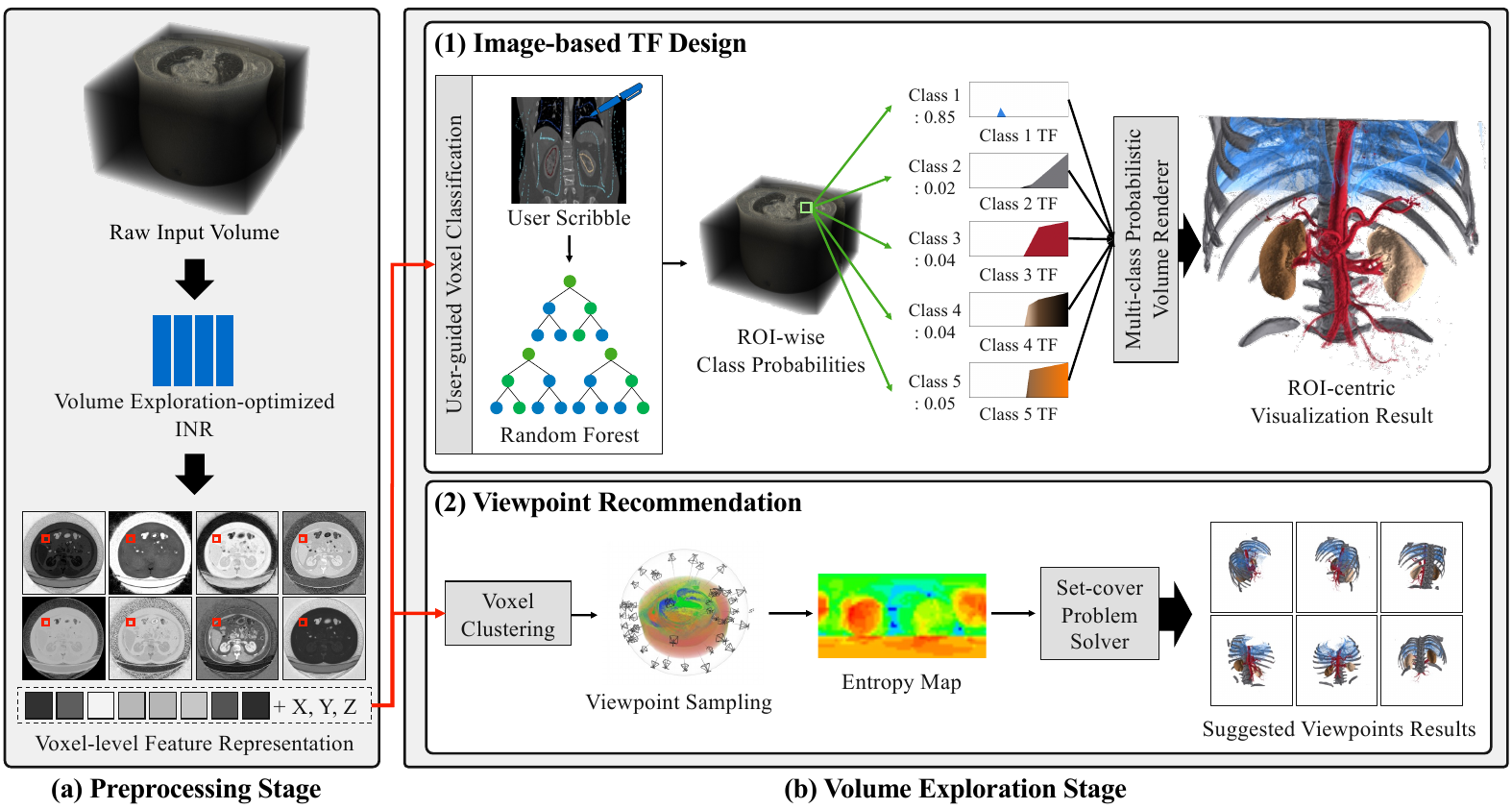}
  \caption{Overview of our NeuVolEx approach.}
  \label{fig:fig2}
\end{figure*}

\section{Related Work}
\subsection{Feature Representation for Volume Exploration}
\subsubsection{Transfer Function Design}

Feature representations in TF design have been used to interactively classify and visualize ROIs. TF design evolved from traditional intensity-based 1D TFs to multi-dimensional TFs that incorporated additional features derived from the volume to provide information beyond intensity. These features served as additional dimensions in TF parameter spaces and acted as indicators that guided users in distinguishing ROIs, which enabled more effective volume exploration. Kindlmann et al. \cite{kindlmann1998semi} pioneered 2D TFs that combined intensity with first- or second-order derivatives to better differentiate ROIs. Other studies identified ROIs using local statistics, texture, or size \cite{max2002optical, cai1995rendering, caban2008texture, correa2008size}. The increasing number of features required more extensive, unintuitive user interaction within widget-based multi-dimensional TF spaces, which in practice constrained the number of features that could be effectively used.

Image-based TF design emerged as an alternative that enabled more intuitive ROI classification while allowing the use of multi-dimensional feature representation. These approaches allowed users to indicate ROIs through interactions on volumes and the extracted feature representations were used to classify ROIs. Tzeng et al. \cite{tzeng2005intelligent} used a feature representation composed of intensity, gradient magnitude, neighboring intensities, and 3D position coordinates. Soundararajan et al. \cite{soundararajan2015learning} further extended this method by introducing probabilistic classifiers that supported uncertainty-based visualization. These studies enabled ROI classification with richer combinations of voxel features. However, because they are still based on local voxel features derived from raw volumes, their limited ability to capture broader geometrical patterns or spatial correlations often makes coherent ROI classification difficult.

Recent studies have explored deep neural networks to learn richer feature representations that could mitigate these limitations. Cheng et al. \cite{cheng2018deep} used a convolutional neural network to extract implicit deep features, but this method relied on extensive training on large-scale external datasets with ground truth labels. Quan et al. \cite{quan2017intelligent}, instead, proposed 3D convolutional sparse-coded feature representations directly learned from an input volume. However, the resulting implicit feature representations are typically high-dimensional (75D), which makes robust performance difficult under practical user supervision budgets. In contrast, Engel et al. \cite{engel2024leveraging} introduced the use of pretrained large-scale feature representations from 2D natural image datasets. This direction offers more generalizable features, but the much higher dimensionality of such representations (387D) makes effective adaptation or fine-tuning to a target volume particularly important for volume exploration. From this perspective, NeuVolEx focuses on optimizing feature representations directly for a given volume, and can therefore be complementary to pretrained-feature-based approaches.

\subsubsection{Viewpoint Recommendation}
Viewpoint recommendation has traditionally relied on hand-crafted feature representations and evaluation metrics to identify how effectively viewpoints convey ROIs within volumes. Takahashi et al. \cite{takahashi2005feature} decomposed a volume into clusters and estimated a global optimal viewpoint based on opacity-TF-based weighting. Bordoloi and Shen \cite{bordoloi2005view} evaluated viewpoints using an information-theoretic metric based on TF-conditioned voxel visibility and noteworthiness. Tao et al. \cite{tao2009structure} incorporated shape and detail features through bilateral-filter-based decomposition and gradient-derived cues and introduced a similarity voting metric. Some approaches instead used feature representations derived directly from rendered images. Ji and Shen \cite{ji2006dynamic} proposed image-based quality measures derived from opacity, color, and curvature distributions in rendered images. Yang et al. \cite{yang2019deep} also measured viewpoint preferences from rendered images and corresponding scores. In contrast, Zheng et al. \cite{zheng2011iview} proposed iView, which used voxel-level features composed of intensity, gradient magnitude, and 3D spatial coordinates from raw volumes to recommend compact and complementary viewpoints that contain ROIs within ROIs. Overall, despite variations in evaluation metrics, prior viewpoint recommendation methods still relied on local feature representation. More expressive features may need to be investigated for robust performance.

\subsection{INR in DVR Visualization}
INRs have emerged as a powerful new paradigm for DVR visualization. In particular, their ability to represent volumetric scalar fields as continuous functions has made them a promising solution for volume compression. Lu et al. \cite{lu2021compressive} first demonstrated the feasibility of representing volumetric scalar fields with coordinate-based multi-layer perceptron (MLP), although their high training and inference costs limited their use in practical applications. To make INR-based volume visualization more practical, Wu et al. \cite{wu2023interactive} proposed a multi-resolution hash encoding approach that enables efficient volume compression and rendering while maintaining visual quality. Wu et al. \cite{wu2024distributed} extended this line of work by improving scalability for large-scale simulations through a distributed architecture. Yang et al. \cite{yang2025meta} accelerated convergence for time-varying datasets by employing meta-learning algorithms to derive effective initialization weights. INRs have also been adopted in expressing TF applied volumes instead of original raw data through the adoption of neural radiance field (NeRF). This coordinate-optical mapping allowed users to manipulate appearance of visible ROIs within DVR images. Yao et al. \cite{yaovisnerf} introduced ViSNeRF, which extends NeRF to time-varying volume series. Tang et al. \cite{tangstylerf} augmented this coordinate-optical mapping with style transfer. As such, the potential of INRs for fundamental volume exploration tasks, such as image-based TF design or viewpoint recommendation, remains largely overlooked. Moreover, prior work has primarily focused on INR outputs, while the feature representations learned during INR training have received limited attention.

\section{Neural Volume Exploration}
\subsection{Overview}
An overview of the NeuVolEx approach for volume exploration is shown in \cref{fig:fig2} using an abdomen CT volume. The proposed volume exploration-optimized INR first learns a voxel-level feature representation (see \cref{fig:fig2}(a)) as a common representation for subsequent volume exploration tasks. This pre-processing stage is performed only once and can be completed within practical time. There are two volume exploration tasks (see \cref{fig:fig2}(b)): (i) For image-based TF design, a user indicates ROIs through simple scribble-based interaction on image slices of the volume. The scribble guidance, together with the voxel-level feature representation, are fed into Random Forest algorithm to estimate the probability of each voxel belonging to each ROI class. The multi-class probabilistic volume renderer then generates ROI-centric visualization results according to user-defined class-wise TFs. (ii) For viewpoint recommendation, the voxel-level feature representation is used to group voxels into ROI clusters. Candidate viewpoints are uniformly sampled on a unit sphere and then analyzed to generate entropy maps that quantify how effectively each viewpoint reveals ROI clusters. A set-cover problem solver selects a set of complementary viewpoints that collectively visualize different ROI clusters.

\subsection{Volume Exploration-Optimized INR}
\begin{figure}[t]
  \centering
  \includegraphics[width=\linewidth]{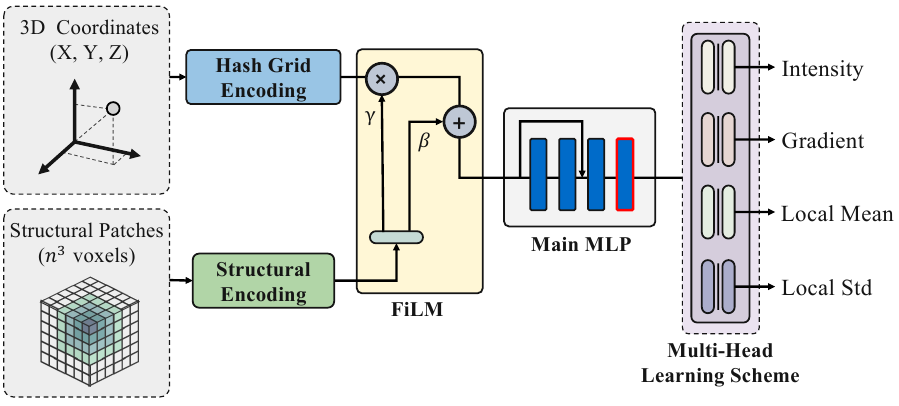}
  \caption{An architectural overview of our volume exploration-optimized INR.}
  \label{fig:fig3}
\end{figure}

\Cref{fig:fig3} shows an architectural overview of our volume exploration-optimized INR. It develops a dual-pathway design to process two complementary inputs. The first pathway produces a positional representation by feeding 3D coordinates into a multiresolution hash grid encoding \cite{muller2022instanthash}. This encoding maps each voxel to trainable features at multiple spatial resolutions. Another pathway samples a local patch of \(n^3\) voxels (where \(n=5\)) centered at each coordinate from the raw volume. This patch is processed by a structural encoder with two hidden layers of 32 channels to generate a structural representation. By encoding local neighborhood around each voxel, this pathway provides local structural context that improves spatial coherence. 

Rather than simply concatenating the two feature representations, we propose a feature-wise linear modulation (FiLM) to effectively combine them, motivated by Perez et al. \cite{perez2018film}. FiLM enables one feature representation to condition another so that the latter reflects additional context. In our case, the structural representation conditions the positional representation based on neighboring voxels around each coordinate. Thus, the positional representation is enriched with local structural context, which improves the separation of neighboring regions with similar intensities and helps preserve fine boundaries. Specifically, the structural representation is passed through a linear layer to generate the channel-wise scale \(\gamma\) and shift \(\beta\), which modulates the positional representation as follows:

\begin{equation}
F_{mod}=F_{pos} \odot (\gamma+1)+\beta
\end{equation}

where \(F_{pos}\) denotes the positional representation derived from the hash grid encoding. An offset of 1 is applied to the scale parameter to maintain an identity mapping at initialization and improve numerical stability. The resulting modulated feature is then fed into the main MLP, which comprises four hidden layers with 64 channels. We further utilize a residual connection between the modulated input and the deeper layers to facilitate gradient flow and stabilize learning on complex volumes.

\begin{figure}[t]
  \centering
  \includegraphics[width=\linewidth]{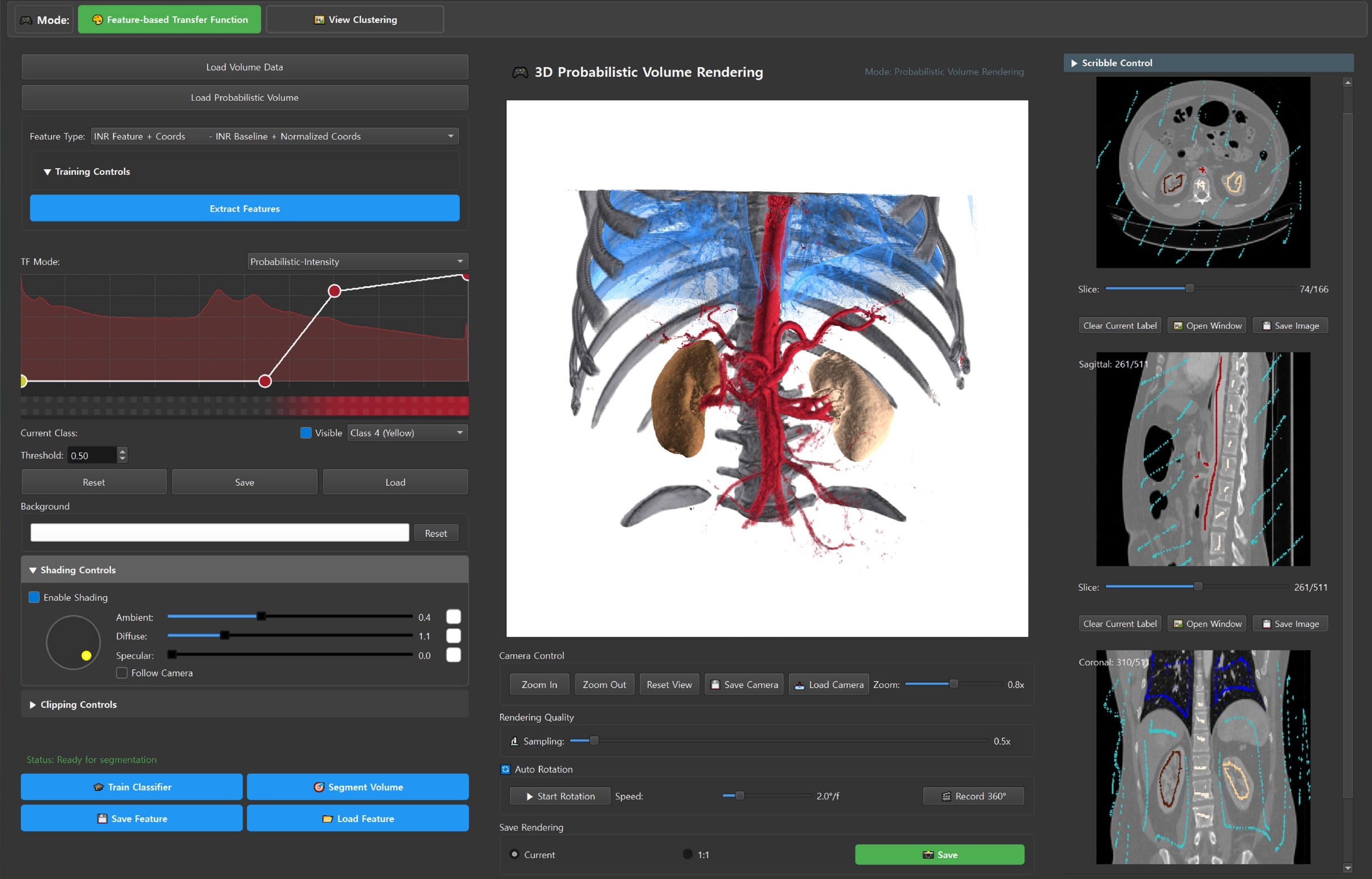}
  \caption{User interface of NeuVolEx for image-based TF design.}
  \label{fig:fig4}
\end{figure}

We further introduce a multi-task learning scheme for training the volume exploration-optimized INR. We note that coherent regions in volumes are characterized not only by intensity values, but also by subtle boundary cues, local homogeneity, and partial volume effects. To capture these properties, the main MLP jointly estimates four targets: intensity, gradient, local mean, and local standard deviation. The total loss \(L_{total}\) is defined as follows:

\begin{equation}
L_{\mathrm{total}} = L_{\mathrm{inten}} + \lambda_{\mathrm{grad}} L_{\mathrm{grad}} + \lambda_{\mathrm{stat}} \left( L_{\mu} + L_{\sigma} \right)
\end{equation}

where \(\lambda_{grad}=1.0\) and \(\lambda_{stat}=0.5\) are empirically determined coefficients. The intensity loss  \(L_{inten}\) preserves the fidelity of the input volume. The gradient loss \(L_{grad}\) encourages the feature representation to be sensitive to structural boundaries that are important for separating adjacent regions. The local statistical losses \(L_\mu\) and \(L_\sigma\) impose local constraints that improve coherent feature responses within the same region. All loss terms are implemented as mean squared error loss. After training, we use the activations from the final hidden layer of the main MLP at each coordinate as a voxel-level feature representation for the downstream volume exploration tasks.

\subsection{Volume Exploration Tasks}
\subsubsection{\textbf{Image-based TF Design}}
\begin{figure}[t]
  \centering
  \includegraphics[width=\linewidth]{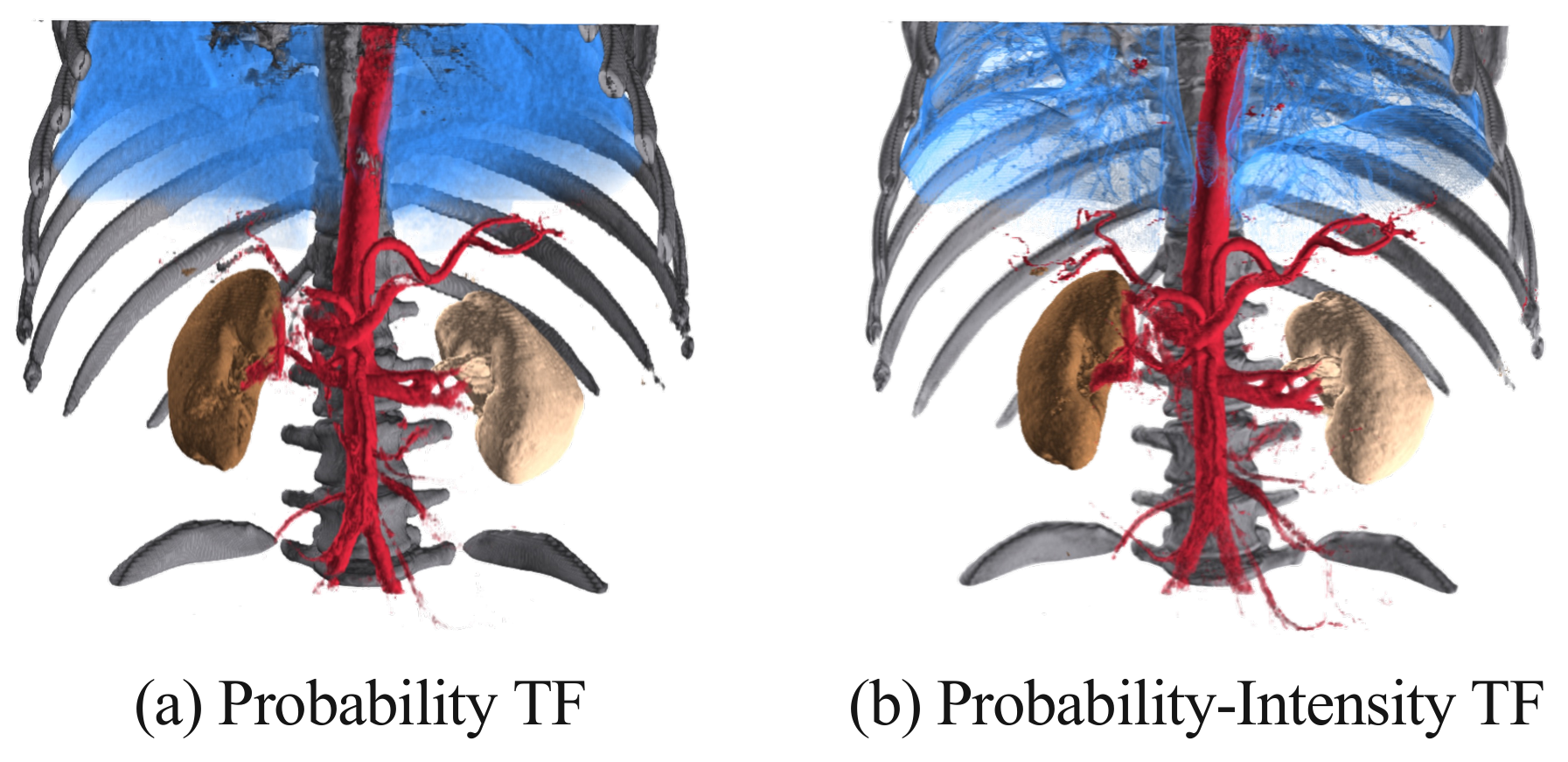}
  \caption{Visualization results using two variants of the multi-class probabilistic volume renderer: the probabilistic renderer and the probability-intensity renderer.}
  \label{fig:fig5}
\end{figure}

NeuVolEx allows users to draw scribbles on cross-sectional image slices of a volume to indicate ROIs to be classified in image-based TF design. \Cref{fig:fig4} illustrates this user interaction using the brush tool in NeuVolEx. A classifier is then trained using the feature representations extracted from the labeled voxels. We choose the Random Forest algorithm due to its robust classification performance and fast training time even under sparse user supervision. The trained classifier predicts ROI-wise class probabilities for the remaining unlabeled voxels, denoted as \(P(v) = [p_1(v), p_2(v), \cdots, p_N(v)]\), where \(p_n(v)\) represents the probability that voxel \(v\) belongs to the \(n\)-th ROI class. These probability vectors are then fed into a standard front-to-back volume compositing pipeline with a multi-class probabilistic volume renderer. We consider two renderer variations as shown in \cref{fig:fig5}.

The probabilistic renderer assigns the optical properties of voxels according to their predicted ROI-wise class probabilities \(P(v)\). The probabilistic renderer calculates the final optical properties through a weighted sum of all \(N\) ROI classes, as follows:

\begin{equation}
C(v) = \sum_n TF_{c,n}\bigl(p_n(v)\bigr)\cdot p_n(v) 
\end{equation}

\begin{equation}
\alpha(v) = \sum_n TF_{\alpha,n}\bigl(p_n(v)\bigr)\cdot p_n(v)
\end{equation}

where \(TF_{c,n}\) and \(TF_{\alpha,n}\) denote the TFs that map the probability of \(n\)-th ROI class to color and opacity, respectively. This approach is particularly useful for domain users who may be less familiar with TF design. It allows users to assign the optical properties intuitively based on how confident each voxel belongs to a given ROI class. However, because intensity variation within the same ROI is not considered, the probabilistic renderer may suppress fine details, such as small vessels or lung bronchus, as shown in \cref{fig:fig5}(a)).

Probability-intensity renderer incorporates voxel intensity \(I(v)\) into TF design to assign color and opacity. In this formulation, the predicted ROI-wise probabilities \(P(v)\) serve as visibility weights that control how much each ROI class contributes to the final voxel visualization. A predefined threshold \(\tau_n\) is applied to suppress low-confidence contributions and retain only ROI classes with probabilities above the threshold. The final color and opacity are calculated as follows:

\begin{equation}
C(v) = \frac{1}{W(v)} \sum_{n:\, p_n(v) \geq \tau_n} TF_{c,n}\bigl(I(v)\bigr)\cdot p_n(v)
\end{equation}
\begin{equation}
\alpha(v) = \frac{1}{W(v)} \sum_{n:\, p_n(v) \geq \tau_n} TF_{\alpha,n}\bigl(I(v)\bigr)\cdot p_n(v)
\end{equation}

where \(W(v) = \sum_{p_n(v) \geq \tau_n} p_n(v)\) is the total weight of ROI-wise probabilities that pass the threshold and is used to normalize the final color and opacity. This formulation allows expert users to perform finer TF adjustments to visualize subtle structures, e.g., small vessels or lung bronchus, as in \cref{fig:fig5}(b).

\subsubsection{\textbf{Viewpoint Recommendation}}
We adopt a method, iView \cite{zheng2011iview}, but use our feature representation, rather than relying only on five explicit local-features, to approximate ROIs within a volume without user supervision. The feature representations of all voxels are clustered into 50 groups using K-means, with each cluster being treated as an ROI. We then uniformly sample 1800 candidate viewpoints on a unit sphere. For each viewpoint, we determine whether each cluster is visible based on the angular relationship between the viewing direction and the representative normal of that cluster. These cluster-wise visibility results are then aggregated into an entropy-based informativeness map, which favors viewpoints that cover a diverse set of visible clusters rather than a few dominant ones. A greedy selection algorithm iteratively selects the viewpoint with the highest remaining informativeness score, thereby producing a compact set of complementary viewpoints.

\begin{table}[tb]
  \caption{Quantitative comparison of ROI classification performance for NeuVolEx against the explicit local-feature method \cite{soundararajan2015learning} and the implicit convolutional-feature method \cite{quan2017intelligent}. The number in parentheses denotes the number of ROIs. For datasets with multiple ROI classes, the results indicate the mean and standard deviation over the ROI classes, whereas for MRA-Brain, only the mean is reported because it contains only one ROI class.}
  \label{tab:Table 1}
  \scriptsize
  \centering
  \setlength{\tabcolsep}{4pt}
  \begin{tabular*}{\columnwidth}{@{\extracolsep{\fill}}ccccc}
    \toprule
    & \multicolumn{4}{c}{F1-score $\uparrow$} \\
    \cmidrule(lr){2-5}
    Method & \makecell{CT-Abdomen\\(5)} 
    & \makecell{CT-Engine \\(7)} 
    & \makecell{CT-Tooth\\(3)}
    & \makecell{MRA-Brain\\(1)}\\
    \midrule
    Local-Feature
    & \makecell{0.8001\\$\pm$0.0780}
    & \makecell{0.7876\\$\pm$0.0758}
    & \makecell{0.8388\\$\pm$0.1511}
    & 0.4335 \\

    Convolutional-Feature
    & \makecell{0.6800\\$\pm$0.1560}
    & \makecell{0.7621\\$\pm$0.1275}
    & \makecell{0.8271\\$\pm$0.1699} 
    & 0.2792\\

    \textbf{NeuVolEx}
    & \makecell{\textbf{0.8302}\\\textbf{$\pm$0.0655}}
    & \makecell{\textbf{0.8368}\\\textbf{$\pm$0.0987}}
    & \makecell{\textbf{0.8795}\\\textbf{$\pm$0.0950}} 
    & \textbf{0.6034} \\
    \bottomrule
  \end{tabular*}
\end{table}

\section{Results}

\subsection{Experiment Settings}
We conducted comprehensive experiments to validate the capabilities of NeuVolEx on two volume exploration tasks quantitatively and qualitatively. For image-based TF design, we measured F1 scores to assess ROI classification performance across a total of 16 distinct ROIs from four different volume datasets. NeuVolEx was compared against two existing methods based on: (i) the explicit local-features proposed by Soundararajan et al. \cite{soundararajan2015learning} and (ii) the implicit-convolutional features proposed by Quan et al. \cite{quan2017intelligent}. We also evaluated the robustness of ROI classification under four variations of user scribbles. Qualitative visualization comparisons are conducted on additional four volume datasets by comparing NeuVolEx against the two image-based TF design methods, as well as the conventional intensity-TF widget \cite{1DTF_robert}. We then further evaluated NeuVolEx on viewpoint recommendation task using three volumes by quantitatively analyzing how well the selected viewpoints cover different ROIs.

We carefully selected ten volumes that represent a range of challenges commonly encountered in DVR visualization practice. Among them, four volumes with ground truth (GT) masks were used for the quantitative evaluation of ROI classification accuracy. A CT-Abdomen volume \cite{dataset_abdomen} with four ROIs representing bone, left kidney, right kidney, and aorta was included because overlapping intensity range makes ROI classification challenging. An MRA-Brain volume \cite{dataset_brain}, which focuses on a single ROI of the cerebral arteries, poses the challenge of reliably differentiating fine tubular structures from background noise. A CT-Tooth volume \cite{dataset_tooth_engine} with the dentin, enamel, and pulp ROIs was included because it has been widely used as a benchmark dataset in many previous publications \cite{quan2017intelligent, diffdvr, illcontext}. An industrial CT-Engine volume \cite{dataset_tooth_engine} that comprises seven internal mechanical components was used to evaluate ROI classification among rigid objects that share similar intensity ranges. GT masks for the CT-Abdomen and MRA-Brain datasets were provided by their respective public repositories. We performed manual GT annotation for the CT-Tooth and CT-Engine datasets due to the lack of publicly available GT masks. Three highly experienced researchers participated in the manual annotation, and any conflicts were resolved by consensus.  

We further used three additional volume datasets with intricate geometries and complex internal organization for the qualitative evaluation of image-based TF design. A CTA-Head volume \cite{dataset_head_cardiac} with the bone, lung, contrast-enhanced vessels, and unenhanced vessels was included to assess how well contrast-enhanced vessels can be distinguished from the others. A CT-Bonsai volume \cite{dataset_bonsai} with tree branches, leaves, dirt, and pot was selected to assess performance on data with dense fine-scale substructures and high geometric complexity. An MR-Kiwi volume with five ROIs representing peel, outer pericarp, inner pericarp, seeds, and core was chosen for testing the ROI classification for the seeds and core, which exhibit subtle intensity variations. We used the remaining three volumes to evaluate NeuVolEx performance on viewpoint recommendation. A Cube synthetic volume, which introduced a slight asymmetry by the engraved text “Vis2026” on one surface, served as a straightforward benchmark. A CTA-Cardiac volume \cite{dataset_head_cardiac} that contains multiple densely arranged and partially occluded internal organs was used to assess the capability of NeuVolEx to identify complex ROI relationships. A Tornado simulation volume that represents elongated flow structures and spatially varying vortex patterns was included to assess whether NeuVolEx could capture these dynamic phenomena from different viewpoints. 

All experiments were conducted on a desktop workstation equipped with an Intel Core i7-14700K CPU, 64 GB of RAM, and an NVIDIA RTX 4090 GPU with 24 GB of VRAM. The operating system was 64-bit Ubuntu 24.04. In the image-based TF design task, we implemented NeuVolEx and the explicit local feature-based method in PyTorch 2.8; the implicit convolutional feature-based method operated on MATLAB R2025b, as in the official repository \cite{dataset_kiwi}. The NeuVolEx was trained as a preprocessing step using the Adam optimizer with a learning rate of 1e-4 for only 100 epochs. The feature training settings for the two comparison methods were optimized for each experiment to obtain the best possible results, in accordance with the published works or experimentally. In the cases where the implicit convolutional feature-based method exceeded memory limits on high-resolution volumes, a server equipped with an Intel Xeon 6540P CPU and 512 GB of RAM was used. We configured the Random Forest classifier with 1000 trees, a minimum sample split of 8, and a square root maximum features criterion without bootstrapping. We set all class thresholds, $\tau$, to 0.5. Voxels with probability values less than 0.5 were considered background and excluded from both the classification metric calculation and subsequent DVR visualization. The multi-class probabilistic volume renderer was implemented using VTK 9.5.0. All visualization outcomes were generated in a probability-intensity renderer with identical TF settings to ensure fair comparisons and visual quality. By contrast, the viewpoint recommendation results were visualized using a standard intensity-TF widget to eliminate any affects of scribble-guided TF design.

\begin{table}[tb]
  \caption{Quantitative comparison of scribble sensitivity for NeuVolEx against the explicit local-feature method \cite{soundararajan2015learning} and the implicit convolutional-feature method \cite{quan2017intelligent} on CT-Tooth across four scribble levels (S1 to S4). The number in parentheses denotes the number of scribble points; the volume contains 1,558,802 voxels.}
  \label{tab:Table 2}
  \scriptsize
  \centering
  \renewcommand{\arraystretch}{1.1}
  \begin{tabular*}{\columnwidth}{@{\extracolsep{\fill}}ccccc}
    \toprule
    & \multicolumn{4}{c}{F1-score $\uparrow$} \\
    \cmidrule(lr){2-5}
    Method & S1 (756) & S2 (1,912) & S3 (9,434) & S4 (29,529) \\
    \midrule
    Local-Feature & \makecell{0.8247\\$\pm$0.1166} & \makecell{0.8388\\$\pm$0.1511} & \makecell{0.8776\\$\pm$0.1129} & \makecell{0.8900\\$\pm$0.1065} \\
    Convolutional-Feature  & \makecell{0.7899\\$\pm$0.1927} & \makecell{0.8271\\$\pm$0.1699} & \makecell{0.8396\\$\pm$0.1565} & \makecell{0.8641\\$\pm$0.1287} \\
    \textbf{NeuVolEx} & \makecell{\textbf{0.8648}\\\textbf{$\pm$0.1053}} & \makecell{\textbf{0.8795}\\\textbf{$\pm$0.0950}} & \makecell{\textbf{0.8911}\\\textbf{$\pm$0.0967}} & \makecell{\textbf{0.8946}\\\textbf{$\pm$0.1105}} \\
    \bottomrule
  \end{tabular*}
\end{table}

\subsection{Image-based TF Design}
\subsubsection{ROI Classification Accuracy}

\begin{figure*}[t]
  \centering
    \includegraphics[width=\textwidth,height=0.75\textheight,keepaspectratio]{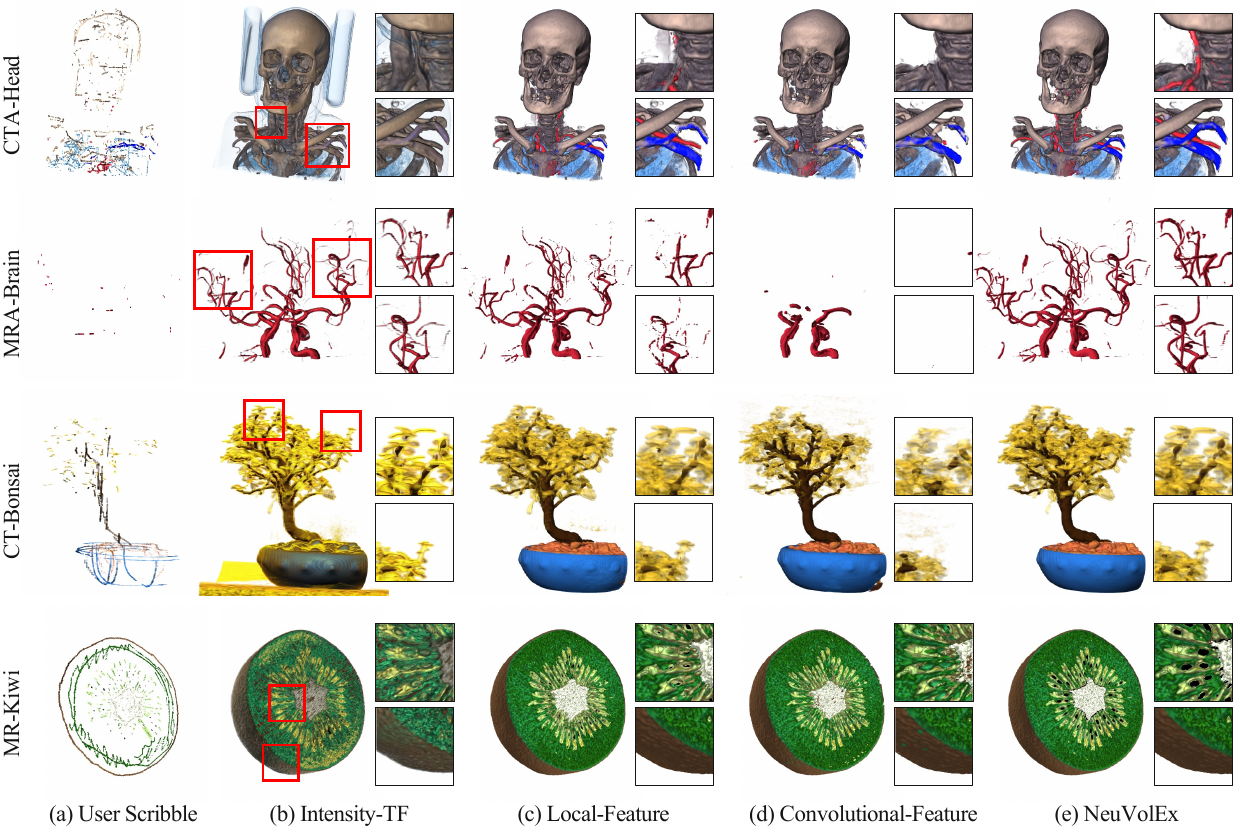}
    \caption{Qualitative visualization comparison on four volumes among NeuVolEx, the intensity-TF \cite{1DTF_robert}, the explicit local-feature method \cite{soundararajan2015learning}, and the implicit convolutional-feature method \cite{quan2017intelligent}, using sparse user scribbles.}
  \label{fig:fig6}
\end{figure*}

\Cref{tab:Table 1} presents the quantitative comparison of ROI classification performance for our NeuVolEx and the two existing methods \cite{soundararajan2015learning, quan2017intelligent} across four different volume datasets. Overall, NeuVolEx consistently achieved the highest mean F1-scores across all volumes, with average improvements of 0.0725 over the explicit local-feature method \cite{soundararajan2015learning} and 0.1504 over the implicit convolutional-feature method \cite{quan2017intelligent}. The three CT volumes exhibited a consistent pattern in which NeuVolEx outperformed both comparison methods by noticeable margins. On CT-Tooth, NeuVolEx achieved the highest F1 score of 0.8795, which outperformed the local-feature method by 0.0407 and the convolutional-feature method by 0.0524. CT-Abdomen yielded the lowest F1-score among all CT volumes, indicating its higher classification difficulty; nevertheless, NeuVolEx still achieved meaningful improvements of 0.0301 and 0.1502 over the local-feature method and convolutional-feature comparison method, respectively. MRA-Brain was the most challenging case, with all methods producing their lowest F1-scores. Despite this difficulty, NeuVolEx maintained an F1-score of 0.6034 and its performance advantage became more pronounced, with margins of 0.1699 over the local-feature method and 0.3242 over the convolutional-feature method. In addition, NeuVolEx showed the lowest standard deviations across all volumes, except for CT-Engine, where the difference from the best local-feature method was only 0.0229.

\Cref{tab:Table 2} presents the quantitative comparison of scribble sensitivity for NeuVolEx and the two comparison methods \cite{soundararajan2015learning, quan2017intelligent} using the CT-Tooth volume. We used four different scribble sets from a sparse level (S1 with 756 points) to the highest level (S4 with 29,529 points), with progressively increasing numbers of labeled voxels. NeuVolEx consistently achieved the best performance across all scribble configurations and, even under S1, already reached a level comparable to the comparison methods under S3 or S4. By contrast, they showed substantial performance degradation under S1, which resulted in a margin of 0.0401 over the explicit local-feature method and 0.0749 over the implicit convolutional-feature method when compared to NeuVolEx. Although the performance gap narrowed as more user supervision (i.e., scribbles) was provided, NeuVolEx still maintained a noticeable advantage, including a margin of 0.0305 over the implicit convolutional-feature method even under the highest level. In addition, NeuVolEx exhibited the lowest standard deviations across all scribble levels, except for S4, where the difference from the explicit local-feature method was negligible (0.0040).

\begin{figure}[tb]
  \centering
  \includegraphics[width=\linewidth]{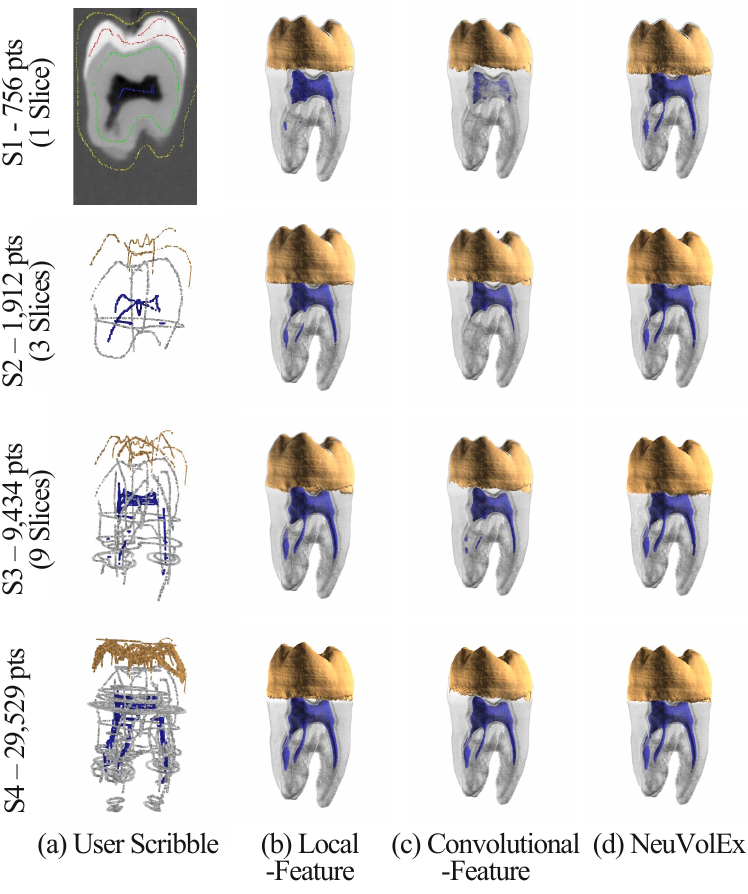}
  \caption{Qualitative visualization comparison of scribble sensitivity for NeuVolEx, the explicit local-feature method \cite{soundararajan2015learning}, and the implicit convolutional-feature method \cite{quan2017intelligent} under four varying scribble levels (S1 – S4) on CT-Tooth.}
  \label{fig:fig7}
\end{figure}

\subsubsection{Visualization Quality}

\Cref{fig:fig6} presents the qualitative comparison of ROI classification and visualization for NeuVolEx against three existing methods \cite{1DTF_robert, soundararajan2015learning, quan2017intelligent} across four different volumes. The user scribbles drawn on the 2D slices of the volumes were visualized in 3D and color-coded according to their corresponding ROI classes. The background class was excluded from the scribble visualizations. NeuVolEx consistently outperformed all comparison methods in terms of ROI classification and visualization across all evaluation datasets. For CTA-Head, which contains five ROIs, NeuVolEx represented them faithfully; in particular, it preserved continuous vessel trajectories, including fine branches, while clearly distinguishing the contrast-enhanced (blue) and unenhanced (red) vessels, as well as separating both from the bones (gray). In contrast, the intensity-TF \cite{1DTF_robert} method failed to differentiate the three ROIs. Although the explicit local-feature method \cite{soundararajan2015learning} enabled better vessel-bone separation, it still suffered from evident discontinuities in the fine vessel branches. Compared with the explicit local-feature method, the implicit convolutional-feature method \cite{quan2017intelligent} showed more severe degradation in vessel connectivity and in distinguishing between the two vessel ROIs.

The visualization results on MRA-Brain, which represented the most challenging case in terms of vessel connectivity (red), again confirmed the superiority of NeuVolEx. In comparison, both local-feature method \cite{soundararajan2015learning} and convolutional-feature method \cite{quan2017intelligent} showed clear degradation in arterial completeness and peripheral branch connectivity. The convolutional-feature method even failed to recover several major vessels. Such visualization enhancement trend from NeuVolEx was consistently observed in non-medical datasets. For CT-Bonsai, which contained delicate tree stems (brown) and leaves (yellow), NeuVolEx faithfully classified and visualized them. Both local-feature method and convolutional-feature method, however, failed to accurately distinguish between the two adjacent, delicate structures. In addition, there was noticeable noise, which became more severe in the convolutional-feature method. Lastly, on MR-Kiwi, in which the seeds (black) and the central core (white) served as the challenging ROIs, NeuVolEx visualized these fine, internal structures with high clarity while maintaining coherent boundaries in the surrounding inner pericarp (light green). The local-feature method, in contrast, largely failed to classify the seeds, which was most pronounced in the convolutional-feature method. Additionally, some misclassification of the outer skin (brown) was observed in the convolutional-feature method. Collectively, these visualization results suggested the robustness of NeuVolEx.

\begin{figure*}[ht!]
  \centering
  \includegraphics[width=\linewidth]{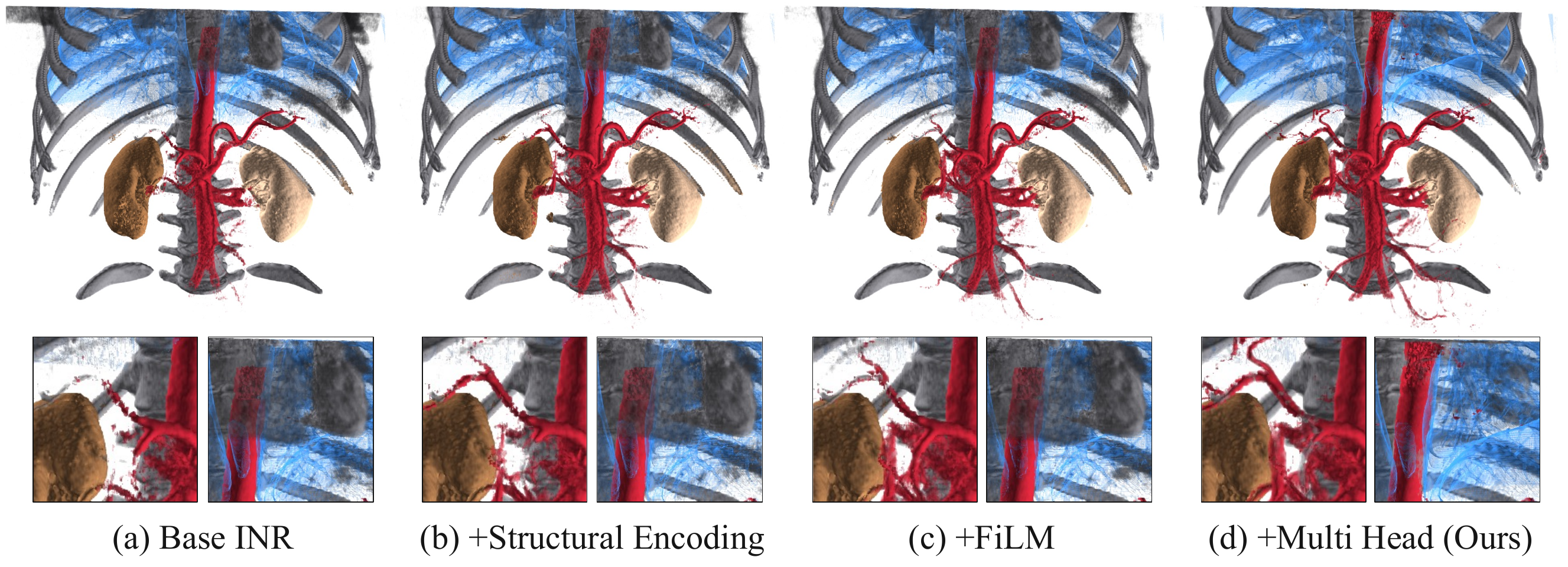}
  \caption{Qualitative ablation study of NeuVolEx on CT-Abdomen.}
  \label{fig:fig8}
\end{figure*}

\Cref{fig:fig7} presents a qualitative visualization comparison of scribble sensitivity among NeuVolEx, the explicit local-feature method \cite{soundararajan2015learning}, the implicit convolutional-feature method \cite{quan2017intelligent} under four varying scribble levels (S1 – S4) on CT-Tooth. CT-Tooth consisted of dentin (white), enamel (yellow), and pulp (blue). The user scribbles were identical to those used in Table 2. NeuVolEx precisely classified and visualized all three ROIs, particularly the continuous and well-connected pulp, even with the extremely limited scribbles from a single image slice (S1 with 756 points). This amount corresponded to only 0.05$\%$ of the total 1,558,802 voxels. The local-feature method, in contrast, produced incomplete visualizations of the lower part of the pulp, which became more pronounced in the convolutional-feature method. The missing visual details gradually recovered as user scribbles increased to S3 (9,434 points across nine image slices). However, the pulp still remained largely disconnected in the convolutional-feature method, which eventually achieved comparable visual quality under the highest level (S4, with 29,529 points), which was approximately 39 times larger than S1 in terms of the number of the scribbled voxels.

\Cref{fig:fig8} presents the visualization results for the ablation study of NeuVolEx on CT-Abdomen volume, which contains five complex ROIs. We evaluated four architectural configurations starting from the base INR model to assess the contribution that each module provided. The base INR model classified and visualized all ROIs to some extent. However, it showed a nontrivial loss of structural integrity, such as in the arterial branch (red), and introduced unwanted misclassification of the heart (gray). The addition of the structural encoding module partially improved the structural integrity by restoring the shape of the aorta which had been fragmented in the base INR model. The subsequent integration of FiLM acted as a complementary module that further restored the vascular details. Finally, the adoption of the multi-task learning module maximized the overall performance by not only recovering all vascular structures, but also suppressing the unwanted heart.

\subsection{Viewpoint Recommendation}
\begin{figure}[tb]
  \centering
  \includegraphics[width=\linewidth]{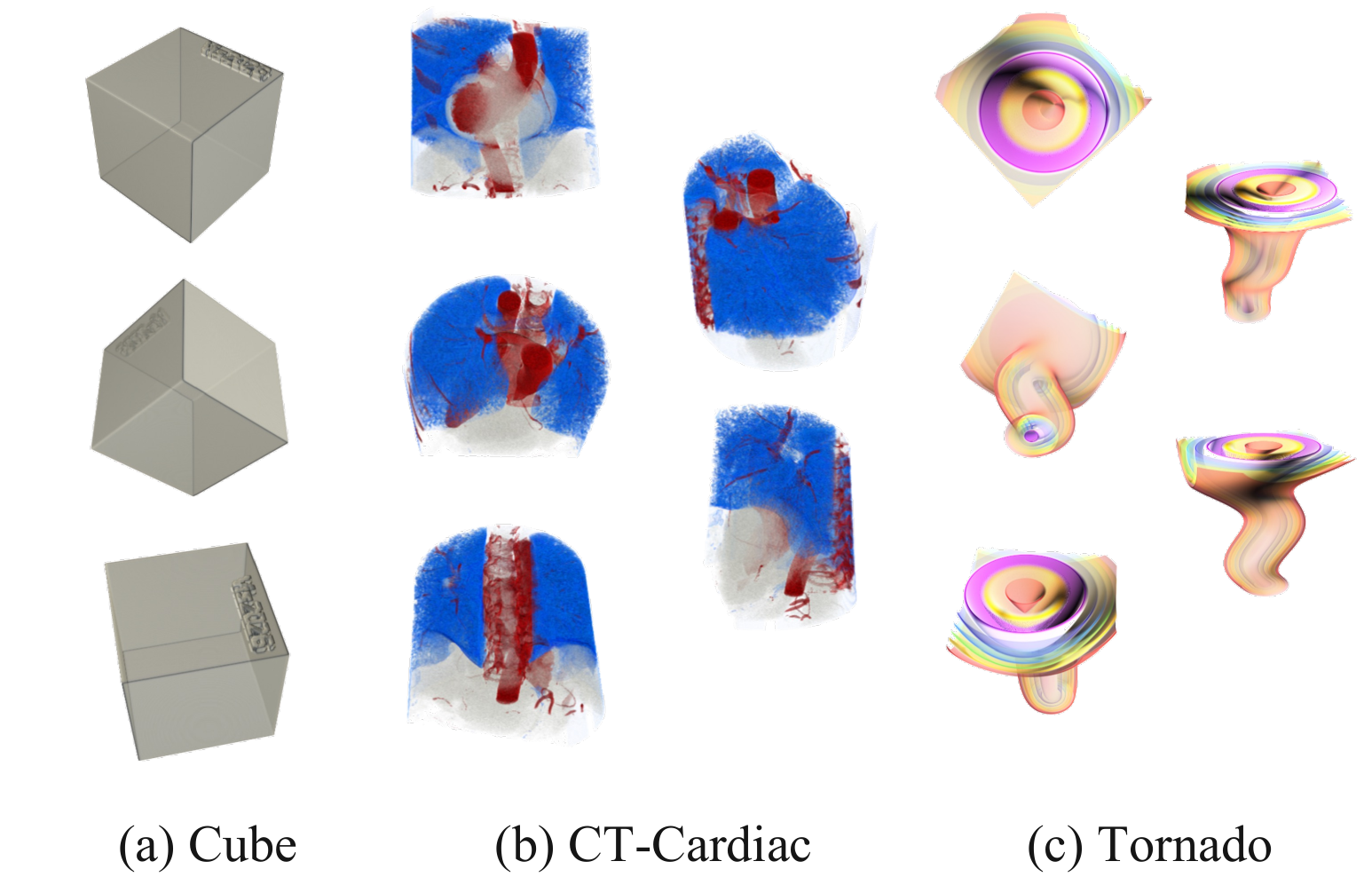}
  \caption{Viewpoint recommendation results from NeuVolEx with three volumes}
  \label{fig:fig9}
\end{figure}

\Cref{fig:fig9} presents the viewpoint recommendation results of NeuVolEx on three different volumes that possess intricate geometries and complex internal organization. The goal of viewpoint recommendation was to automatically produce a compact set of complementary viewpoints. Since this task does not admit a quantitative ground-truth answer, we assessed NeuVolEx based on whether it could recommend a set of viewpoints that covered major ROIs or contained ROIs occluded in other viewpoints, while avoiding redundant observations. For Cube, the suggested viewpoints consisted of two corner perspectives and one diagonal perspective. The two corner viewpoints revealed three faces simultaneously from opposite vertices, which effectively covered the overall geometry with minimal overlap. The additional diagonal viewpoint uniquely revealed the fine topographical details of the engraved surface text. Similarly, in CT-Cardiac, five distinct viewpoints comprehensively spanned the front, back, top, and lateral perspectives of the cardiac model. Together, these viewpoints visualized vessel ROIs that were partially hidden from individual viewpoints. Lastly, for Tornado, five selected viewpoints illustrated the complex vortex morphology of the layered ROIs. A top-down viewpoint directly revealed the vortex center and its concentric layer organization, while lateral viewpoints clearly exposed the vertical curvature and the twisting funnel-shaped pattern.

\subsection{Computational Performance}

\begin{table}[tb]
  \caption{Computational performance of NeuVolEx on the ten evaluation volume datasets with varying resolutions and experiment settings. The bottom three volumes were used for viewpoint recommendation and therefore have no classifier training results.}
  \label{tab:Table 3}
  \centering
  \scriptsize
  \renewcommand{\arraystretch}{1.1}
  \resizebox{\columnwidth}{!}{%
  \begin{tabular}{ccccccc}
    \toprule
      &
      & \multicolumn{2}{c}{Feature Training}
      & \multicolumn{3}{c}{Classifier Training} \\
    \cmidrule(lr){3-4} \cmidrule(lr){5-7}
        Dataset
      & \makecell{Volume\\Resolution}
      & \makecell{Memory\\Usage (GB)}
      & \makecell{Processing\\Time (s)}
      & \makecell{Number of\\Scribbles (pts)}
      & \makecell{Number\\of ROIs}
      & \makecell{Processing\\Time (s)} \\
    \midrule
    CT-Tooth     & $103 \times 94 \times 161$  & 1.85  & 11.82  & 1912  & 3 & 1.75 \\
    CT-Bonsai    & $256 \times 256 \times 77$  & 2.07  & 36.93  & 6373  & 4 & 2.96 \\
    CT-Engine    & $256 \times 256 \times 128$ & 2.50  & 61.26  & 5975  & 7 & 5.96 \\
    MRA-Brain    & $332 \times 432 \times 100$ & 2.97  & 106.30 & 14257 & 1 & 2.46 \\
    CTA-Head     & $256 \times 256 \times 230$ & 3.04  & 112.15 & 8363  & 4 & 5.47 \\
    MR-Kiwi      & $256 \times 256 \times 256$ & 3.28  & 125.07 & 7257  & 5 & 8.96 \\
    CT-Abdomen   & $512 \times 512 \times 167$ & 7.29  & 331.50 & 19358 & 5 & 21.84 \\
    \midrule
    CTA-Cardiac  & $512 \times 512 \times 452$ & 18.75 & 904.36 & --    & -- & -- \\
    Cube         & $256 \times 256 \times 256$ & 3.28  & 125.21 & --    & -- & -- \\
    Tornado      & $256 \times 256 \times 256$ & 3.28  & 125.14 & --    & -- & -- \\
    \bottomrule
  \end{tabular}%
  }
\end{table}

\Cref{tab:Table 3} shows the computational performance of NeuVolEx across the ten evaluation volumes with various resolutions and experiment settings. Feature training of NeuVolEx, as a preprocessing step, showed the computational costs that scaled proportionally with volume resolution. The smallest volume, CT-Tooth, used less than 2GB memory and required around 12 seconds, which can be regarded as nearly real-time preprocessing. Even in the largest CTA-Cardiac, the feature representation was obtained in around 15 minutes with acceptable memory overhead. For the top seven volumes, which were used for image-based TF design, we confirmed that the classifier training (with user interaction) remained efficient regardless of all experimental settings.

\section{Discussion}
The results collectively indicate that the main strength of NeuVolEx lies in its INR-based feature representation, which (i) robustly characterizes ROIs within volumes (see \cref{tab:Table 1} and \cref{fig:fig8}); (ii) enhances two fundamental volume exploration tasks, image-based TF design and viewpoint recommendation, across a variety of volume datasets (see \cref{fig:teaser}, \cref{fig:fig6} and \cref{fig:fig9}); and (iii) supports practical usability in interactive workflows (see \cref{tab:Table 3}).

The quantitative ROI classification results (see \cref{tab:Table 1}) show that NeuVolEx can robustly discriminate between ROIs within volumes. We attribute this capability largely to the use of the INR-based feature representation learned during volume training. In the DVR community, INRs have mainly been used to compactly and implicitly represent volumes for reconstruction purposes. This implies that INR can also produce a feature representation that encodes voxel-specific characteristics. In this work, we focus on this intermediate feature representation,  which has received limited attention in the DVR community, where prior works have primarily treated INRs as coordinate-based functions for accurate volume reconstruction. 

Although the base INR provides feature representations useful for ROI discrimination (see \cref{fig:fig8}(a)), these features are not inherently tailored for volume exploration. The base INR is primarily trained to approximate individual voxels precisely, whereas volume exploration requires spatially coherent grouping of adjacent voxels with subtle variations while still separating them from different local neighborhoods. To better capture this spatial coherence, NeuVolEx incorporates a structural encoder and a multi-task learning scheme. The structural encoder captures structural integrity (see \cref{fig:fig8}(b)), while the multi-task learning scheme helps refine ambiguous boundaries and intricate shapes (see \cref{fig:fig8}(c)). The ablation results (see \cref{fig:fig8}(d)) show that these modules play complementary roles, thus enabling NeuVolEx to better adopt INR-based feature representations to  volume exploration. 

TF design is a fundamental volume exploration task, which depends on user-guided classification based on feature representations. The results (see \cref{fig:fig6}(e)) show that NeuVolEx can accurately classify and visualize ROIs only with simple scribble inputs on image slices. This capability becomes clearer when compared with prior image-based TF methods (see \cref{fig:fig5}). The method proposed by Soundararajan et al. \cite{soundararajan2015learning} uses explicit local-features derived from volumes, but its limited capability to capture broader spatial coherence often leads to missing fine structures or misclassifying adjacent regions, as seen in the discontinuous vessel trajectories in CTA-Head or the poor separation between seeds and core in MR-Kiwi (see \cref{fig:fig6}(c)). In contrast, the method proposed by Quan et al. \cite{quan2017intelligent} introduces implicit convolutional-features to encode spatial context and hierarchical relationships. However, its inherent high-dimensional feature representation requires extensive and delicate scribble inputs to achieve accurate ROI classification (see \cref{fig:fig7}(c)). Under a practical interaction budget, this method unfortunately produces degraded results, for example missing substantial portion of the arterial network in MRA-Brain (see \cref{fig:fig6}(d)). NeuVolEx alleviates both limitations, thereby enabling more accurate classification of fine details and small regions.

Viewpoint recommendation is another important volume exploration task, which relies on clustering voxels based on their feature representations without user supervision. The results (see \cref{fig:fig9}) show that NeuVolEx can robustly identify a compact set of complementary viewpoints that collectively expose diverse ROI clusters across different volumes. This suggests that the INR-based feature representation generalizes beyond user-guided TF design and can also support unsupervised viewpoint recommendation. 

Practical volume exploration requires high usability, and in our experiments, it depends on both interaction efficiency and computational efficiency. The TF design results (see \cref{fig:fig7} and \cref{tab:Table 2}) show that NeuVolEx achieves accurate ROI classification using only 756 scribble inputs from a single image slice, which corresponds to only a tiny fraction of the full volume (103 x 94 x 161). In contrast, the convolutional-feature comparison method \cite{quan2017intelligent} requires much denser interaction across multiple image slices to obtain comparable results, e.g., 29,529 points, which is 39 times more than NeuVolEx. The usability also depends on whether feature representations can be trained within practical computational budgets. The results (see \cref{tab:Table 3}) show that the feature training cost of NeuVolEx scales approximately linearly with volume resolution; even for a large volume of 512×512×452, it requires only around 15 minutes and 18.75 GB of VRAM. A direct GPU-based comparison with the convolutional-feature method is not possible because its publicly available GPU implementation is not currently executable, but our supplementary CPU-based experiment suggests that the convolutional-feature method requires significantly greater memory resources and impractical training times (more than a week) for the same volume. These interaction and computational advantages largely stem from the compact INR-based feature representation of NeuVolEx, in contrast to the high-dimensional convolutional-features used in the comparison method. 

The overall results demonstrated that NeuVolEx is applicable across a variety of volumes with different imaging modalities and ROI complexities, as well as diverse volume exploration conditions. This generalizability motivates extending NeuVolEx to other types of volume datasets, such as multimodal volumes and time-varying volume series. It may also be extended to other volume exploration tasks, such as identifying clipping planes that reveal ROIs or establishing correspondence between 3D volumetric structures and 2D visualizations. Such extensions will likely require additional adaptation, since NeuVolEx is not directly applicable to these new problems in its current form. Future work will therefore focus on adapting and optimizing NeuVolEx for broader tasks and data generalization. Another important direction is to further improve scalability. Although NeuVolEx already shows acceptable computation times and memory requirements for large volumes, continued advances in modern simulations and imaging hardware are rapidly increasing data scale. Handling such scales will require further optimization of NeuVolEx for more efficient training and deployment.

\section{Conclusion}

We introduced NeuVolEx, a novel approach that extends the role of INRs from their conventional use in volume reconstruction to volume exploration. The results demonstrate that our INR-based feature representation, optimized with a structural encoder and a multi-task learning scheme, improves ROI classification and clustering, thereby supporting two fundamental volume exploration tasks: image-based TF design and viewpoint recommendation. Together with its high usability, these findings suggest that NeuVolEx can serve as a robust and practical framework for volume exploration in DVR.

\bibliographystyle{abbrv-doi-hyperref}

\bibliography{template}

\appendix 
\crefalias{section}{appendix} 

\end{document}